\documentclass[aps, nofootinbib, preprint, showpacs]{revtex4}

\usepackage{amssymb}
\usepackage{amsbsy}
\usepackage{amsmath}

\newcommand {\mtext}[1] {\quad \mbox{#1} \quad}
\newcommand {\eqn}  [1] {equation~(\ref{#1})}

\newcommand {\nn}   {\nonumber}
\newcommand {\be}   {\begin{equation}}
\newcommand {\ee}   {\end{equation}}
\newcommand {\ba}   {\begin{array}}
\newcommand {\ea}   {\end{array}}
\newcommand {\bea}  {\begin{eqnarray}}
\newcommand {\eea}  {\end{eqnarray}}

\newcommand {\half} {\textstyle\frac{1}{2}\,}

\newcommand {\Ein}  {Ein\-stein}
\newcommand {\Sch}  {Schwarz\-schild}

\newcommand {\GHS}  {Gar\-fink\-le, Horo\-witz and Strom\-inger}
\newcommand {\Nord} {Nord\-str\"om}
\newcommand {\NN}   {New\-ton--\Nord}

\newcommand {\Per}  {Pe\-rel\-man}
\newcommand {\Jor}  {Jor\-dan}

\newcommand {\BD}   {Brans--Dicke}
\newcommand {\JBD}  {\Jor--\BD}

\newcommand {\Poin} {Poin\-ca\-r\'e}
\newcommand {\PC}   {\Poin~con\-jec\-ture}

\newcommand {\VOL}  {\mbox{$\boldsymbol \omega$}}
\newcommand {\DIV}  {\mbox{div}}

\newcommand {\EG}   {\Ein~gra\-vi\-ty}
\newcommand {\SEG}  {standard \Ein~gra\-vi\-ty}
\newcommand {\RF}   {Ricci flow}
\newcommand {\RFG}  {\RF~gravity}

\newcommand {\vp}   {vo\-lu\-me--pre\-ser\-ving}
\newcommand {\VP}   {vo\-lu\-me--pre\-ser\-va\-tion}
\newcommand {\EMT}  {ener\-gy--mo\-men\-tum tensor}
\newcommand {\EOM}  {equa\-tion of mo\-tion}
\newcommand {\LELST}{low--ener\-gy li\-mit of string theo\-ry}
\newcommand {\ST}   {sca\-lar--ten\-sor}
\newcommand {\Lap}  {\Delta\,}

\newcommand {\p}    {\partial}
\newcommand {\cont} {\! \cdot \!}
\newcommand {\Pm}   {P_m}
\newcommand {\Qm}   {Q_m}

\newcommand {\where}{\mtext{where}}
\newcommand {\const}{\mbox{const}}
\newcommand {\und}  {\mtext{and}}

\newcommand {\ie}   {i.e.}
\newcommand {\eg}   {e.g.}
\newcommand {\rhs}  {r.h.s.}

\begin{document}

\title{Ricci Flow Gravity}

\author {Wolfgang Graf \email{wolfgang.graf@univie.ac.at}}

\address {Institut f\"ur theoretische Physik, Universit\"at Wien, Austria}

\email{wolfgang.graf@univie.ac.at}

\begin{abstract}

A theory of gravitation is proposed, modeled after the notion
of a {\em Ricci flow}.
In addition to the metric an {\em independent volume} enters as
a fundamental geometric structure.
Einstein gravity is included as a limiting case.
Despite being a scalar--tensor theory the coupling to matter
is different from Jordan--Brans--Dicke gravity.
In particular there is no adjustable coupling constant.
For the solar system the effects of
Ricci flow gravity cannot be distinguished from Einstein gravity
and therefore it passes all classical tests.
However for cosmology significant deviations
from standard Einstein cosmology will appear.  
\pacs{04.50.+h, 04.20.Cv, 02.40.Hw}

\end{abstract}

\maketitle

\section{Introduction}

A generalization of \Ein's theory of gravity is developed.
It has a purely geometric foundation, including in addition to a metric
an {\em independent volume}. Although related
to \ST~theories \`a la \JBD~and to string cosmology,
it is nevertheless different:
the basic field equations are in the form of {\em \RF~equations}, generalized to include matter.
\Ein's theory is included as the limiting case of no flow.
The volume scalar has two interpretations: geometrically, it is responsible
for \VP~and physically it obeys a mass--zero real scalar wave equation.
This is also the main difference with \JBD~theories, where the scalar couples to the trace
of the \EMT.
As a consequence, in general the \EMT~is not
anymore ``conservative'' in the ordinary sense
of $\nabla\cont T = 0$, and point particles do not move anymore on geodesics,
having a Newton--\Nord~potential.
But from Noether's fundamental relation conservation still follows from symmetry.
Nevertheless, in ``ordinary'' solar--system and astrophysical settings,
the corrections are negligible: the theory
cannot be distinguished from \Ein's and therefore passes all the standard tests.
However in a cosmological setting, deviations from \SEG~are
to be expected. This will be the subject of a forthcoming paper.

We will proceed as follows:
after this brief introduction, in section \ref{motivation}
the motivations for this kind of extension are discussed.
In section \ref{volman} a short introduction to volumetric
manifolds is given, emphasizing the notion of \VP~in section \ref{volpres}.
Before establishing the definitive field equations of \RFG~in section \ref{rfg},
several other choices are discussed in section \ref{feqns},
with emphasis on the main differences with respect to \Ein's
and in particular to \JBD~theories.
Section \ref{physint} refers to the physical interpretation of the volume scalar in \RFG.
Finally, in section \ref{compliance} the viability of the theory
with respect to the standard tests is discussed.
The conclusion in section \ref{conclusion} ends this paper.

\section{Motivation and Inputs}
\label{motivation}

The present work is principally motivated by the conviction
that the notion of ``volume'' has an existence independent
from any metric --- in fact, it must be considered
to be a {\em pre--metric} concept.
Curiously, such an independent volume had
not been taken into consideration in physical theories
until relatively recently.
Even in differential geometry it is almost ignored. Perhaps the reason
for this neglect is that in most circumstances
there is a {\em canonical} volume element, based
on other geometric structures considered to be
more basic. For example, in Riemannian geometry
the volume element density is defined in terms of the metric.
In particular, the important operation of {\em Hodge dual}
for differential forms is conventionally based on such a Riemannian volume element.

However, from the gravitational sector of the \LELST~(\ie, compactification to dimension $n=4$) there
comes the suggestive hint (cmp.~Garfinkle, Horowitz and Strominger \cite{GHS91})
that when both a dilaton scalar
and a two--form are present, the dilaton scalar enters
the expression for the volume element density
when defining a ``natural'' Hodge dual operator.\footnote
{for the dilaton general concept, see Sundrum \cite{Sun03}; for the connection of gravity to strings,
see Ort\'in \cite{Ort04}}
This was taken as the starting point to develop
a theory of {\em geometric dilaton gravity} (Graf \cite{Gra03}).
Although the particular coupling does not exactly correspond
to the coupling suggested by string theory,
wormholelike solutions were obtained.

Recently a breakthrough on \RF~methods was achieved by
\Per~\cite{Per02, Per03a, Per03b}, developing the decisive tools to solve the famous \PC~on
the topological characterization of the three--sphere.
Based on 3--d (compact and positive--definite)
Riemann spaces, smoothly deformed
by a {\em \RF}~(RF), the ``basic'' RF equation was originally defined by
\be
\p_t\,g_{ik} = -2\,R_{ik},
\ee
where $R_{ik}$ is the Ricci--tensor corresponding to a ``time--dependent'' three--metric $g_{ik}$.
Also a special class of diffeomorphisms was considered, with vector $\vec v$
which is essentially the gradient of a scalar $\phi$ in the sense of $v^i = g^{ik}\,\p_k\phi$.
The so generalized RF equation then becomes
\be
\label{de_turck_eqn}
\p_t\,g_{ik} = -2\,(R_{ik} + \nabla_i\nabla_k\,\phi).
\ee
Although such equations have already been studied since the early eighties starting with the seminal works of 
Hamilton \cite{Ham82} and DeTurck \cite{DeT83}, an essential insight of \Per~was 
to recognize that the \rhs~of this equation\footnote
{the expression in parenthesis in the \rhs~of \eqn{de_turck_eqn}
has its own measure--theoretic meaning and is also known
under the name ``Bakry--\'Emery'' tensor (cmp.~Lott \cite{Lot02})}
can be expressed as the {\em gradient} of an appropriate functional.
This functional involves a ``measure'' given in terms of an {\em independent volume element}.
The gradient property allows to apply a series of standard analytical tools.
And the introduction of the measure gives an extra flexibility,
analogous to a choice of gauge.\footnote
{in particular he envisages {\em \vp~flows}
and certain diffeomorphic images thereof}

Since the works of Hamilton and DeTurck just mentioned,
geometric flows have been applied to a variety of geometric,
topological and analytical problems.\footnote
{see the recent monograph of Chow and Knopf \cite{ChK04} on Ricci flows
(not covering \Per's contributions),
the introduction by Topping \cite{Top05},
and the lecture notes by Morgan and Tian \cite{MoT06}}
Flow--like equations are also not unfamiliar to physicists,
the earliest and most well--known being the renormalization group equations
in quantum field theory (for an introduction, see Mitter \cite{Mit05}), and also the nonlinear $\sigma$--model (\eg, Lott \cite{Lot86}, 
Oliynyk, Suneeta and Woolgar \cite{OSW06}, Tseytlin \cite{Tse07}).
After Ellis \cite{Ell84} called attention to the cosmological ``fitting problem'', 
the usefulness of the \RF~to deal with volume--averaged inhomogeneities was 
immediately recognized and continues to be an active area of research 
(\eg, Carfora and Marzuoli \cite{CaM84}, Buchert and Carfora \cite{BuC02}, and the recent review by Buchert \cite{Buc07}).
An overview of flow techniques in physics is given in Bakas \cite{Bak05}.

Another motivation came however with the insight,
that the basic equations derived from the \LELST~can
be put into a form suprisingly similar to \RF~equations
when besides the metric only a dilaton scalar is kept.
The main formal difference is the number of dimensions
and the signature of the corresponding Riemann spaces:
whereas the ``classic'' RF equations refer to a parameter--dependent
truly Riemannian three space evolved by an extrinsic ``time'' parameter,
the reformulated string theory
equations refer to a four--dimensional Lorentzian spacetime,
which is evolved along the directions of an intrinsic
vector field.

Neither the ``classic'' RF approach nor string theory
suggest any hints about the coupling of geometry to external matter fields.
Therefore we will spend some time
to prepare the field in order to include
other external matter.
As not only geodesy of the motion of ``test particles'' will in
general be violated, but also ``conservation''
(in the sense of $\nabla\cont T = 0$),
we will be especially careful to lay a
coherent and stringent foundation.
The Noether identities will be our main guide.
As result we will get \RFG~(RFG).

For the history of \ST~theories and their current status, we refer to
Brans \cite{Bra05}, and to the recent monographs
of Fujii and Maeda \cite{FuM03} and Faraoni \cite{Far04}.

\section{Volumetrical Manifolds}
\label{volman}

In Graf \cite{Gra03} we already introduced the notion
of a volume manifold and its specialization for the case a nondegenerate
metric exists. Let us briefly recapitulate the main notions.
First, we introduced the fundamental concept of a {\em volume structure},
which has to be considered as independent from any metric.
This is just a non--negative $n$--form density \VOL,
and makes the manifold a {\em volume manifold}.
Secondly, we will need of course a {\em metric structure}.
However, it does not need to be compatible with the volume structure.
This difference is encoded by means of the {\em volume scalar} $\phi$ by
$\VOL = \omega\,e^{-\phi}$,\footnote
{the factor $-1$ of $\phi$ is purely conventional --- here we follow \Per \cite{Per02},
in contrast to string theory, where $-2$ is preferred}
where $\omega := |\det{g}|^{1/2}\;dx^1\wedge dx^2\wedge \cdots \wedge dx^n$
is the usual metrical (\ie, Riemannian) volume element density.
Furthermore, for the metric derivative along a vector $X$ we have
$\nabla_X\,\VOL = -(X \cont \p \phi) \:\VOL$ (the dot denoting a contraction)
as a measure of incompatibility.

Such a manifold, endowed both with
an independent volume and a metric structure,
we will denote by {\bf volumetrical manifold}.
Whereas the manifold is considered to be smooth,
both metric and volume element density are allowed to diverge
or to be degenerate, when they are not locally smooth.

Already in a volume manifold the Gauss~theorem for a vector $\xi$
can be expressed very compactly in terms of differential form densities as
\be
\int_{b\,\Omega} \xi\cont\VOL = \int_\Omega d\;(\xi\cont\VOL),
\ee
where $b\,\Omega$ is a two--sided hypersurface bounding the  n--dimensional region~$\Omega$.
The scalar factor $\DIV\,\xi$ in the relation $d\,(\xi\cont\VOL)=(\DIV\,\xi)\,\VOL$
is also better known under the name of {\em divergence} of the vector $\xi$.
Evidently the div--operator only depends on the particular choice of $\VOL$
and not on any metric.

\section{Volume Preserving Lie Flows}
\label{volpres}

In a differentiable manifold, the thing coming closest
to an autonomous first order differential equation for a ``vector'' $x(t)$,
\be
\dot x = f(x),
\ee
is the notion of a {\em Lie equation}
\be
\pounds_\xi \,F = G,
\ee
where $F$ and $G$ are geometric objects (\eg, tensors),
$\xi$ is some vector field and $\pounds_\xi \,F$ denotes
the Lie derivative of $F$ along $\xi$. In the simplest case $\xi$ and
$G$ are considered as given and $F$ to be determined.
However in the applications we have in mind,
{\em all} elements of the equation will be dynamically determined,
$G$ depending nonlinearly on $F$ and its partial derivatives,
and even $\xi$ will become dynamical.

In the theory of ordinary differential equations,
such systems of first--order equations which guarantee the long--term existence
both to the future and the past, are also called {\em flows}
and can be characterized by the {\em one--parameter Abelian group property}
of their solutions.
As well--known, Lie operators share
exactly the same one--parameter Abelian group property (at least locally)
by means of the exponential map. We can therefore speak of a {\em Lie flow}.

In a volume manifold, a Lie flow with vector $\xi$
is called {\em \vp} (or VP)\footnote
{this is a local concept in contrast to the much weaker global definition of Huisken \cite{Hui87}} 
if
\be
\pounds_\xi \,\VOL = 0, \mtext{or equivalently,} \DIV\,\xi=0.
\ee
In the rest of this paper we will try to make plausible a particular scalar--tensor extension
of \Ein~gravity in terms of a {\em \vp~Ricci Lie flow}
in a volumetric manifold.

\section{A choice of Scalar--Tensor Field Equations}
\label{feqns}

Assuming that the total Lagrangian (or at least the field equations) can
be uniquely split into a pure geometrical part and the ``physical'' part,
we can already draw important conclusions about both the algebraic
and the differential properties of the ``physical'' \EMT~just
from examining the purely geometrical part.
Note that whereas \Per's analysis is ``metric--centered'',
with an auxiliary scalar, in the following physical applications
this scalar will play a role at the same conceptual level as the metric.
Therefore the ``classical'' Lagrangian approach is appropriate.

Let us start with the ``geometrical'' Lagrangian living on a volumetric manifold $M$,
\be
\mathfrak L = \VOL\,(R+\lambda\,(\nabla\phi)^2),
\ee
where $\VOL := e^{-\phi}\,\omega$ and $(\nabla\phi)^2 := g^{ij}\,\p_i\phi\:\p_j\phi$
and $\lambda$ is a constant parameter.
Despite its simple form it not only includes the one used initially by \Per~and
in the \LELST~(when ignoring the axion and the other moduli fields),
but which also is essentially the \JBD~Lagrangian.

Defining the {\em volume factor} $\Phi := e^{-\phi}$,
we then have as variational derivatives
(up to volume element, $g$--dualizations of $P$ and a common sign $-1$)
\bea
\label{p_eq}
\frac{\delta\,\mathfrak L}{\delta g_{ik}} \sim P_{ik} &:=& G_{ik} -
\Phi^{-1}\left(\nabla_i\nabla_k - g_{ik}\;\Delta \right) \Phi \nn \\
&& + {~}
\lambda\,\Phi^{-2} \left(\nabla_i\Phi\nabla_k\Phi - {\textstyle\frac{1}{2}}\,g_{ik}\,(\nabla\Phi)^2 \right),  \\
\label{q_eq}
\frac{\delta\;\mathfrak L}{\delta \,\phi} \sim Q &:=&
R - 2 \,{\lambda}\,\Phi^{-1} \Lap\Phi + \lambda\,\Phi^{-2} (\nabla\Phi)^2 ,
\eea
where $G$ denotes the \Ein~tensor $G_{ik} := R_{ik}-\half R\,g_{ik}$ and $\Delta := \nabla^2$
the d'Alembertian.
For the above Lagrangian the {\em Noether identity} can be written compactly as
\be
\label{noether_div}
\DIV \, (\tilde P^i_k\,\xi^k) = P^{ik}\,\pounds_\xi\,g_{ik} + Q\,\pounds_\xi\,\phi,
\ee
with some tensor $\tilde P^i_k$ to be determined by it. More conventionally,
\be
\label{noether_conv}
\nabla_i\left(\Phi\,\tilde P^i_k\,\xi^k\right) = \Phi\left(P^{ik}\,\pounds_\xi\,g_{ik} + Q\,\pounds_\xi\,\phi\right).
\ee
As this identity must hold for any smooth vector $\xi$, we get separately
\be
\tilde P_{ik} = 2\,P_{ik} \und \nabla_i(\Phi\, \tilde P^i_k) = Q\,\p_k\,\Phi.
\ee
Note that from \eqn{noether_conv} follows conservation in the proper sense, if $\vec\xi$
is a {\em simultaneous} Killing vector both of the metric and of the  scalar, even if $Q\neq0$.

The following algebraico--differential relations evidently hold:
\begin{description}
\item [{\bf symmetry:}]$\tilde P_{ik}=\tilde P_{ki}$, and
\item [{\bf balance:}] $\nabla_i(\Phi \tilde P^i_k) = Q\,\p_k\,\Phi$.
\end{description}
Up to this point we only made use of identities, but not of any field equations.
In particular if we equate $P$ and $Q$ to their corresponding physical
quantities, these algebraico--differential relations will be ``impressed'' on them.
In fact, it is not even necessary that they follow from a Lagrangian.

But note that there is a dependency not only on $\lambda$
but also on the number $n$ of dimensions,
making $n=4$ and $n=3$ (for $\lambda=1$) somewhat special.
This is most evident in the relation
\be
P + Q = \left((n-1)-2\lambda\right)\,\Phi^{-1} \Lap\Phi - \frac{\lambda}{2}\,(n-4)\,\Phi^{-2} (\nabla\Phi)^2,
\ee
where $P$ is the ``trace'' $P:=g^{ik}\:P_{ik}$.
Assuming from now on $n=4$, this simplifies to
\be
\label{pq_sum_eqn}
P + Q = \big(3-2\,\lambda\big)\,\Phi^{-1} \Lap\Phi.
\ee

Let us define the {\em geometrical \EMT} $P_{ik} := 1/2\,\tilde P_{ik}$
and more closely examine the corresponding balance relation
\be
\nabla_i(\Phi\, P^i_k) = {\textstyle\frac{1}{2}}\,Q\,\partial_k\,\Phi.
\ee
The following cases can be distinguished, when equating the geometrical quantities
$P$ and $Q$ to their ``physical'' counterparts $\Pm$ and $\Qm$:
\begin{description}
\item [{\bf a) Pure Einstein,}] {$\phi=0$}
\item [{\bf b) Conformally Einstein,}] {$\lambda=3/2$}
\item [{\bf c) ``Conservative'',}] {$Q = 0$}
\item [{\bf d) VP flow,}] {$\Lap\Phi=0$}
\item [{\bf e) Fully dynamical.}]
\end{description}
Each of these choices will now be discussed individually.

\subsection{Pure Einstein}

This is just the ``compatibility mode'', or ``Einstein--limit'' $\phi\to0$ (if it exists).
It is thus a volumetric theory only in the trivial sense of $\phi=0$.

\subsection{Conformally Einstein}

The system of equations is underdetermined. This becomes evident by
going to the Einstein--frame by means of the conformal transformation $g'_{ik} = e^{\phi}\,g_{ik}$,
where the scalar field drops out completely.

\subsection{``Conservative'': \JBD}
\label{jbd}

From $Q=0$ there follows ``conservation'' in the usual sense of $\nabla_i(\Phi\, P^i_k) = 0$.
This assumes the particular relation $P\Phi = (3-2\lambda)\,\Lap\Phi$,
which could either be postulated or obtained by a specially tailored Lagrangian.
In order to have a more familiar looking equation, $P_{ik}$ could be equated to the physical
quantity $T_{ik}$ over $P_{ik} = \Phi^{-1} T_{ik}$,
so that in fact ``conservation'' in the sense of $\nabla_i\,T^i_k = 0$ would result.
This kind of ``conservation'' was considered as absolutely essential in the
closely related \ST~theories of \Jor~and \BD~(in short, JBD theories).\footnote
{cmp.~Jordan \cite{Jor55}, Weinberg \cite{Wei72}, part II, ch.~7, \S 3
and Fujii and Maeda \cite{FuM03}}
In fact, this can be achieved as follows:
their scalars $\phi$ (resp.~$\kappa$) must be identified with $\Phi$,
and we must identify their coupling parameters
$-\omega$ (resp.~$\zeta$) with~$\lambda$.
Moreover $\lambda \neq 3/2$ has to be assumed, otherwise the conformally \Ein~theory would result.
Then $1/\phi$ (resp.~$\kappa$) is interpreted as the (variable) gravitational constant.
Both for \Jor's and \BD's material \EMT~it is supposed that it does {\em not} depend on the scalar~$\phi$.
However in \Jor's theory it is the product $\kappa^2\,T^i_k$ which is ``conserved''.\footnote
{this is suggested by his interpretation of the Kaluza--Klein decomposition}
In particular, for a ``dust model'' geodesy of $\vec u$ (resp.~$\kappa^2 \vec u$)
and conservation of $\rho \vec u$ (resp.~$\kappa^2 \rho \vec u$)
still follow when staying in the original conformal (``Jordan'') frame.

\subsection{Volume--preserving Flow}

When not a conformally \Ein~coupling, from \eqn{pq_sum_eqn} the condition $P+Q=0$
is equivalent to $\Lap\Phi=0$, which in turn is equivalent to \VP~$\pounds_\xi\,\VOL = 0$.
This translates to the scalar condition
\be
\Pm+\Qm=0
\ee
for the corresponding ``material'' quantities. Let us call such a coupling to matter
a {\em \vp~material coupling} (VPMC), and assume it to hold troughout this section.
Then
\be
\nabla_i(\Phi\, P^i_k) = -{\textstyle\frac{1}{2}}\:P\,\partial_k\,\Phi.
\ee
Evidently, if the trace $\Pm$ of the \EMT~vanishes the VPMC is satisfied if we set $\Qm=0$.
Then the standard ``conservation'' continues to hold. This is the case \eg~for the Maxwell field.

As an important example where $P\neq0$, let us take the ideal fluid model,
where the material \EMT~is given by $\Pm^{ik} := T^{ik} = \rho\,u^iu^k + p\,\Pi^{ik}$,
and $\Pi^i_k := \delta^i_k+u^iu_k$ is the projector orthogonal
to the trajectory with (normalized) tangent $\vec u$.
Its trace is $T=3p-\rho$. To satisfy the VPMC, we must set $Q_m=-T$.
Specializing to pure dust we get
$\nabla_i(\Phi\, \rho \,u^iu^k) = \half\rho\,g^{ik}\,\partial_i \Phi$.
Splitting into tangential and orthogonal parts, we then get the separate equations
\be
\label{eom}
\nabla_i\,(\rho\, u^i) = -\half \rho\,\dot\phi \und
\dot u^i = \half \Pi^{ik}\,\partial_k\,\phi.
\ee
Due to the nonvanishing of the \rhs~of these equations, both ``conservation of matter'' and 
geodesy for ``test particles'' are broken unless $\phi=\const$.
And due to the particular form of the \EOM~\ref{eom}b (\ie, being proportional to a gradient)
we have in fact got a {\em \NN--term}.\footnote
{recall that around 1912--13 \Nord~developed a precursor relativistic gravitational theory,
where the gravitational potential $\phi$ obeys a
Minkowskian potential equation, $\Lap\phi = 0$.
This was shown in 1914 by \Ein~and Fokker to admit a conformally Minkowskian formulation}

Concerning the divergence expression (\ref{eom}a),
it can nevertheless be rewritten as a {\em conservation law},
$\nabla_i\:(\Phi^{{1/2}}\, \rho u^i) = 0$.
Therefore, for such a theory with \vp~flow,
both the \EOM~as well as the ``conservation of dust matter''
are not anymore the well--known standard expressions from \Ein~or \JBD~theory.
It can be expected that this will have profound consequences in a cosmological setting.

We will continue the discussion of \vp~theories in {section~\ref{rfg}},
where we further specialize to the coupling parameter $\lambda=1$.

\subsection{Fully dynamic Scalar Field}

Here the scalar $\phi$ is dynamically determined
by a set of field equations obtained via a suitable Lagrangian,
and no case of the previously discussed ones fits.
This would normally be the ``standard'' procedere in physics,
where not only the Lagrangian is set up as a linear combination of individual Lagrangians,
each one describing a different matter model,
but in addition possibly introducing some extra ``potential terms'' containing
$\phi$ and $\p\phi$, or even let $\lambda$ depend on $\phi$. 
However this will in general prevent a simple geometrical interpretation in terms of a flow,
and in particular will lack the crucial VP property.
For example, our geometric dilaton gravity (Graf \cite{Gra03}) belongs to this more general class.

\section{Ricci Flow Gravity}
\label{rfg}

The class of \vp~volumetric theories can be further refined
by requiring the particular value $\lambda=1$
of the coupling, as is common in the \LELST.
With this particular value the field equations can be rearranged into
an explicit flow--like form and we get the {\bf \RFG~equations} (RFG equations)
\bea
\pounds_\xi\,g_{ik} &=& 2\,(R_{ik}- \bar T_{ik}), \\
\pounds_\xi\:\VOL &=& 0, 
\eea
describing {\bf Ricci flow gravity} (RF gravity).
Here the {\em flow vector} is defined in terms
of the volume scalar as $\vec\xi = -g^{-1}\partial\phi$,\footnote
{the arbitrary minus--sign is taken in view of cosmological applications}
and $\bar T_{ik} := 8\pi\,(T_{ik} - \half T\,g_{ik})$.\footnote
{we use troughout the sign- and units conventions of Misner, Thorne and Wheeler \cite{MTW73}}
In contrast to the JBD equations, they have a much simpler structure and an immediate
{\em geometric} character.
Through their particular flow--like form, they exhibit a strong
{\em dynamical} touch: broadly speaking, the rate of change of the metric is driven by the
difference of the geometrical and the physical energy momentum tensors.
Evidently, when the flow vector can be ignored (e.g., when it vanishes)
equations equivalent to \Ein's are obtained.
In this sense \EG~is a special case of \RFG.

The RFG vacuum equations are equivalent to JBD's vacuum equations with $\omega=-1$.
More remarkably is the fact that they are also equivalent
to the equations following from the \LELST~when the
standard dilaton coupling with $\lambda=1$ is chosen and besides the metric
only the dilaton scalar is kept.
And of course there is a strong resemblance
to \Per's \RF~equations
which can be made more evident as follows.
Consider tentatively on $M := M_3\times T$ the vector $\vec \xi := -(\p_t + \vec v)$
and the metric $g_{ik}$ with line element $ds^2 = d\sigma^2-dt^2$.
Then the generalized RF \eqn{de_turck_eqn} can be written as $\pounds_{\xi} \, g_{ik} = 2\,R^{(3)}_{ik}$,
which differs (in content, but not in form)
only on the \rhs~from the corresponding RFG vacuum equation.\footnote
{the connection between some solutions of the \RF{} equations for $n=3$ and solutions of the \Ein{} equations for $n=4$ was further elaborated by Bleecker \cite{Ble90} and by List \cite{Lis05}}

This coincidence of seemingly different approaches
could signal a deeper raison d'\^ etre.

\section{On the physical interpretation of the scalar $\Phi$}
\label{physint}

The scalar $\Phi$ was here interpreted geometrically
in the context of a volumetrical manifold
as the volume factor.
In the theories of JBD the corresponding scalar is essentially interpreted
as ``gravitational constant'' $\kappa$ --- more precisely
$\kappa = \Phi$ in Jordan's theory, whereas $\kappa = 1/\Phi$ in \BD~theory.
However this physical interpretation cannot be upheld anymore in a \vp~theory like RFG
where $\kappa$ is constant.

Due to the fact that the volume factor $\Phi$ of a \vp~theory obeys
the d'Alembertian wave equation $\Lap\Phi=0$ it must therefore be interpreted
as a {\em massless real scalar field}.
By the tenets of relativistic quantum mechanics this
corresponds to a {\em totally uncharged massless bosonic particle}.\footnote
{except for a ``dilaton charge''; see the discussion in next section}
The \VP~will also be instrumental to guarantee an almost perfect compliance with the standard solar system
tests of gravity.
This is a fair return for the price we had to pay for giving up the geodesy
of ``test particles''.

\section{Compliance of \RFG~with the Standard Tests}
\label{compliance}

For the standard solar system tests the corresponding
generalization of the \Sch~metric is needed.
The general asymptotically flat and
static spherically symmetric vacuum RFG solution 
with $\Phi \to 1$ for $r\to\infty$ can be written as
\bea
&&ds^2 = -Y^{\gamma-\sigma}\,dt^2 + Y^{-\gamma-\sigma}\left(dr^2 + Z^2\,d\Omega^2\right),\\
&&\Phi \equiv e^{-\phi} = Y^\sigma, \where \\
&&Y := \frac{r-r_+}{r-r_-}, \quad Z^2 := (r-r_+)(r-r_-), \nn
\eea
with $\gamma^2 + \sigma^2 = 1$,
and it is assumed that $r \geq r_+ > r_- \geq 0$.\footnote
{in the ``degenerate'' case $r_+ = r_-$ the metric is locally flat and the volume factor constant}
Being for $r > r_+$ a vacuum RFG solution,
it is also the corresponding general JBD vacuum solution in the ``Jordan''--frame.
But whereas in JBD gravity the source of the volume factor
$\Phi$ for a mass point has to be
a certain nonzero distribution supported by $r=r_+$, 
in RF gravity due to $\Lap\Phi=0$ it must be sourceless.
This can be shown to hold even for a compactly supported smooth static spherically symmetric 
energy--momentum tensor as source, if both metric and volume factor are smooth and the
manifold is simply--connected.
Therefore for RFG $\sigma=0$, whereas for JBD $\sigma = 1/2\,(3+2\,\omega)^{-1}$.

This can also be expressed more conveniently in terms of
the ``dilaton charge'' $D$,
which in the context of the \LELST~is defined for
a static solution with Killing vector $\eta$
(normalized to $\eta^2=-1$ at infinity) as
\be
D = \frac{1}{4\pi} \oint \eta\cont\xi\cont \VOL,
\ee
where the integral is taken over a closed
and externally orientable 2--sphere at spatial infinity.\footnote
{cmp.~\GHS \cite{GHS91}}
For RF gravity the two--form density
$\chi := \eta\cont\xi\cont \VOL$ is even closed, $d\,\chi=0$,
for any {\em stationary} solution with Killing vector $\eta$
so that the above integral only depends on the homology class of
the closed externally orientable 2--sphere.
In particular it vanishes if this 2--sphere bounds.
With the flow vector $\vec \xi = \sigma\,(r_+-r_-)\,Y^{\gamma+\sigma}\,Z^{-2}\,\p_r$
for the above solution this results in $D=\sigma\,(r_+-r_-)$.
For vanishing dilaton charge the \Sch~solution is evidently reobtained
after substituting $r$ by $r+r_-$, setting $m=(r_+-r_-)/2$ and assuming $m > 0$.
Thus for the standard solar--system tests
the flow vector vanishes and we have {\em full compatibility}
with \EG, which passes these tests with ever increasing accuracy
(cf.~Will \cite{Wil06}).\footnote
{to compare, for JBD gravity to pass the current tests $|\omega| > 4\times10^4$ must be assumed}

Of course where the flow vector does not vanish, \RFG~and
\EG~will lead to different answers. Using heuristically the term ``charge''
as introduced above (possibly without stationarity)
we note that differently from the ``mass charge'' $m$,
the ``dilaton charge'' $D$ can have any sign.\footnote
{this allows the dilaton scalar to act ``repulsively'', as shown in Graf \cite{Gra03}}
Therefore the contributions to the total charge of a collection of ``charged regions''
can still sum up to zero, so as to make the
\NN--terms of the \EOM~insignificant for sufficiently big distances.\footnote
{\eg, for a ``multipole charge'' when the distance is
much bigger than the individual ``charges''}
This should be considered to be in fact the case for ``ordinary matter''
building up planets, stars and perhaps, galaxies.
Significant differences are however to be expected in a cosmological setting,
where the ``big bang'' will affect the behaviour of the volume scalar $\phi$.

Although for the ``compliant mode'' $\phi=\const$ evidently it makes no difference
if the metric is interpreted in the geometric frame or in the Einstein frame,
this is not so in the general case
where even the equations of motion for a point particle are modified.
We have to chose the {\em particular conformal frame},
where the field equations find their
``most natural expression''.
This is the {\em geometrical} frame
with an independent volume element density.

\section{Conclusions}
\label{conclusion}

Motivated by the neglect of the notion of an independent volume
and led by the appeal of \Per's approach to solve the \PC,
as well as by the equations following from the \LELST, we developed
the equations of \RFG~as a natural extension of \EG.
The main differences with regard to other \ST~theories
were worked out in the framework of volumetric manifolds.
The \VP~of the flow turned out to be of decisive importance
for the theory and allowed it to essentially agree with
\Ein's under non--cosmological settings and not too small distances
in the case of vanishing ``total dilatonic charge''.

\section*{Acknowledgements}

I thank the members of the relativity group of the University of Vienna
for discussions and useful suggestions, and in particular Peter C.~Aichelburg. 
Partial financial support from the Fundaci\'on Federico is also acknowledged.


\begin{thebibliography}{10}


\bibitem{GHS91}
Garfinkle D, Horowitz GT, Strominger A: \emph{Phys Rev D} 1991, \textbf{43}:3140. 
[Erratum: {\em Phys Rev D} 1992, {\bf 45:}3888].

\bibitem{Sun03}
Sundrum A: arXiv:hep--th/0312212.

\bibitem{Ort04}
Ort\'in T: \emph{Gravity and Strings}. 
Cambridge: Cambridge University Press 2004.

\bibitem{Gra03}
Graf W: \emph{Phys Rev D} 2003, \textbf{67}:024002. [arXiv:gr--qc/0209002].

\bibitem{Per02}
Perelman G: arXiv:math.DG/0211159.

\bibitem{Per03a}
Perelman G: arXiv:math.DG/0303109.

\bibitem{Per03b}
Perelman G: arXiv:math.DG/0307245.

\bibitem{Ham82}
Hamilton RS: \emph{J Diff Geom} 1982, \textbf{17}:255.

\bibitem{DeT83}
DeTurck DM: \emph{J Diff Geom} 1983, \textbf{18}:157.

\bibitem{Lot02}
Lott J: arXiv:math.DG/0211065.

\bibitem{ChK04}
Chow B, Knopf D: \emph{The Ricci Flow: An Introduction}. 
Providence: American Mathematical Society 2004.

\bibitem{Top05}
Topping PM: \emph{Lectures on the Ricci flow}. 
Cambridge: Cambridge University Press 2006.

\bibitem{MoT06}
Morgan J, Tian G: arXiv:math.DG/0607607.

\bibitem{Mit05}
Mitter P: arXiv:math--ph/0505008.

\bibitem{Lot86}
Lott J: \emph{Commun Math Phys} 1986, \textbf{107}:165.

\bibitem{OSW06}
Oliynyk T, Suneeta V, Woolgar E: \emph{Nucl Phys B} 2006, \textbf{739}:441.
[arXiv:hep--th/0510239].

\bibitem{Tse07}
Tseytlin AA: \emph{Phys Rev D} 2007, \textbf{75}:064024.
[arXiv:hep--th/0612296].

\bibitem{Ell84}
Ellis GFR: In \emph{Proceedings of the Tenth International Conference on General Relativity and Gravitation}. 
Edited by Bertotti B, de~Felice F, Pasolini A, Dordrecht: Reidel 1984:215.

\bibitem{CaM84}
Carfora M, Marzuoli A: \emph{Phys Rev Lett} 1984, \textbf{53}:2245.

\bibitem{BuC02}
Buchert T, Carfora M: \emph{Phys Rev Lett} 2002, \textbf{90}:031101.

\bibitem{Buc07}
Buchert T: arXiv:0707.2153 [gr--qc].

\bibitem{Bak05}
Bakas I: arXiv:hep--th/0511057.

\bibitem{Bra05}
Brans C: arXiv:gr--qc/0506063.

\bibitem{FuM03}
Fujii Y, Maeda KI: \emph{The Scalar--Tensor Theory of Gravitation}. 
Cambridge: Cambridge University Press 2003.

\bibitem{Far04}
Faraoni V: \emph{Cosmology in Scalar--Tensor Theory}. 
Dordrecht: Kluwer 2004.

\bibitem{Hui87}
Huisken G: \emph{J reine angew Math} 1987, \textbf{382}:35.

\bibitem{Jor55}
Jordan P: \emph{Schwerkraft und Weltall}. 
Braunschweig: Vieweg 1955.

\bibitem{Wei72}
Weinberg S: \emph{Gravitation and Cosmology: Principles and Applications of the General Theory of Relativity}. 
New York: Wiley 1972.

\bibitem{MTW73}
Misner CW, Thorne KS, Wheeler JA: \emph{Gravitation}. 
San Francisco: Freeman 1973.

\bibitem{Ble90}
Bleecker DD: \emph{J Geom Phys} 1990, \textbf{7}:363.

\bibitem{Lis05}
List B: \emph{PhD thesis}, Freie Universit\"at Berlin 2005.

\bibitem{Wil06}
Will CM: \url{http://www.livingreviews.org/lrr-2006-3}.

\end{thebibliography}
\end{document}